\documentclass[sigconf]{acmart}
\usepackage[T1]{fontenc}
\usepackage[utf8]{inputenc}
\usepackage{textcomp}
\usepackage{xurl}
\usepackage{enumitem}
\SetEnumitemKey{mynosep}{noitemsep, nosep, topsep=0pt, partopsep=0pt, parsep=0pt, itemsep=0pt, leftmargin=*}

\usepackage{xcolor}
\colorlet{light-gray}{gray!20}
\usepackage{newfloat}
\usepackage{listings}
\DeclareFloatingEnvironment[fileext=lst,placement={!htbp},name=Listing]{listing}
\newcommand\enableACMPreprint{
  \settopmatter{printacmref=false}
  \setcopyright{none}
  \renewcommand\footnotetextcopyrightpermission[1]{}
  \pagestyle{plain}
}

\acmConference[IMX-LATAM 2020]{IMX-LATAM 2020: Workshop IMX in Latin America}{June 17, 2020}{Barcelona, Spain}

\enableACMPreprint

\begin{document}

\title{A Brief Overview about D-Profile of Ginga DTV Receivers}

\author{Marcelo F. Moreno}
\affiliation{%
\institution{UFJF}}
\email{moreno@ice.ufjf.br}

\author{Débora C. Muchaluat-Saade}
\affiliation{
\institution{MídiaCom Lab/UFF}}
\email{debora@midiacom.uff.br}

\author{Guido Lemos de Souza Filho}
\affiliation{%
\institution{UFPB}}
\email{guido@ci.ufpb.br}

\author{Raoni Kulesza}
\affiliation{%
\institution{UFPB}}
\email{raoni@ci.ufpb.br}

\author{Álan L. V. Guedes}
\affiliation{%
\institution{TeleMídia/PUC-Rio}}
\email{alan@telemidia.puc-rio.br}

\renewcommand{\shortauthors}{Moreno \textit{et. 
al.}}

\begin{abstract}
The Brazilian DTV system standards have been recently revised in order to address new use cases related to deeper integration between broadcast services and broadband services. 
Such an evolution could not disrupt the current DTV services since Brazil and many ISDB-T countries are still under the switch-off process from analog to digital. 
The middleware layer is the best candidate for such an incremental (yet powerful) evolution. 
In the case of Ginga, the Brazilian DTV middleware, as an open platform based on receiver profiles, the results were impressive. 
In this paper, we focus on the advances and new use cases addressed by the D-profile Ginga DTV receiver specification carried out in the Brazilian DTV Forum under a collaborative effort led by the Academia. 
\end{abstract}

\maketitle

\section{Introduction}

With the widespread adoption of high-speed Internet services, TV devices also use a broadband channel. 
This channel allows the delivery of both linear and non-linear (on-demand) content from the Internet.
A combining scenario that uses both broadband and traditional broadcast channels (terrestrial or cable) may take advantage of strong points from each one. 
The former has advantages regarding live events and high audiences, while the last benefits from better navigability and recommendations.
ITU calls this combination scenario as IBB (Integrated Broadcast-Broadband) service. 

According to ITU\footnote{\url{https://www.itu.int/rec/T-REC-J.205/en}}, an IBB service simultaneously provides an integrated experience of broadcasting and interactivity relating to media content, data, and applications from multiple sources, where the interactivity is sometimes associated with broadcasting programmers.
This channel combination enables IBB to support applications that use \textit{multi-sourced media content}.
For instance, such services can provide delivery content through the broadcast channel, and some other content can be delivered through the broadband channel such as complementary media (\textit{e.g.} short clips and ads) or media extensions (\textit{e.g.} second stereoscopic eye, saleable resolution).
To enable this convergent ecosystem, IBB also supports \textit{API for application integration}.
These applications could be delivered from broadband and broadcast channels, and even from the TV vendor itself textit{e.g.}, from an app. repository) and home area network textit{e.g.}, smartphones). 

Interactive TV applications already existed before IBB initiatives, such as the established standardized Brazilian TV middleware Ginga\cite{itu2014h761}.
Since its conception, Ginga supports developers to use \textit{multi-sourced media content}.
Regarding immersive media, we cite the new support for \textbf{HDR and 4K} video formats.
Recently the Brazilian DTV system standards have been revised in order to address \textit{API for application integration} for new use cases related to deeper integration between broadcast services and broadband services. 
Such an evolution could not disrupt the current DTV services since Brazil and many ISDB-T countries are still under the switch-off process from analog to digital. 
The middleware layer is the best candidate for such an incremental (yet powerful) evolution. 
In the case of Ginga, the Brazilian DTV middleware, as an open platform based on receiver profiles, the results were impressive. 

In this paper, we focus on the advances and uses cases addressed by the D-profile Ginga DTV receiver.
These specification efforts were carried out in the Brazilian DTV Forum under a collaborative effort led by the Academia. 
However, we must go beyond these new specs and rethink interactivity for future television services.

The paper is organized as follows. 
Section \ref{related-work} presents related work.
In Section \ref{architecture}, we introduce Ginga and its architecture.
The next sections discuss advances and new components of the Ginga architecture~(sections \ref{media-delivery}-\ref{ginga-cc-ws}).
Then, we discuss some usage scenarios in Section \ref{usecases}.
Finally, Section \ref{final-remarks} presents our final remarks and future work.

\section{Related work}\label{related-work}

There are three other technologies applied in the same field with similar purposes:  HybridCast, HbbTV and ATSC 3.0.

Hybridcast\footnote{\url{http://www.iptvforum.jp/en/hybridcast}} uses HTML5 and is standardized in Japan as the second generation of multimedia coding specifications that succeeds in the first one, based on BML. 
The specification is completely new, but HybridCast receivers can present BML applications. 
The system offers a variety of services through a combination of broadcast and broadband telecommunication resources and features. 
Features also include basic support for companion devices and synchronization by stream events only.

The HbbTV\footnote{\url{http://www.hbbtv.org}} specification is based on existing standards and web technologies. 
It is a proposal from ETSI that succeeds in its former middleware specification called Multimedia Home Platform (MHP) for interactive digital television. 
HbbTV is a whole new specification that is not compatible with MHP. 
IBB and non-IBB services can coexist in a DTV system that adopts HbbTV by proper signaling. 
HbbTV 2.0 finally added support for some kind of media synchronization, companion devices, and new media types, a set of features that were not supported in previous versions. 
HTML5 became the base language included in HbbTV 2.0. 

ATSC\footnote{\url{http://www.atsc.org/standards/atsc-3-0-standards}}~(Advanced Television System Committee) refers to digital television standards developed in the  US.
ATSC 3.0 has been designed with HTML5 and IP as the key components of IBB  services.
In other words, HTML5 as a  presentation layer technology and IP as an underlying delivery protocol.

\section{Ginga Architecture Overview}
\label{architecture}

\begin{figure}[!ht]
    \centering
    \includegraphics[width=\linewidth]{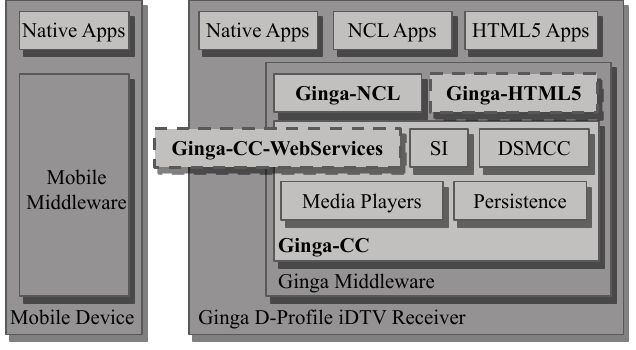}
    \caption{D-Profile Ginga Receiver}
    \label{fig:architecture}
\end{figure}

Ginga was initially proposed for terrestrial DTV systems, then the same architecture and facilities were applied to IPTV. 
Its modular architecture also allows for use with other transport systems \textit{e.g.}, IPTV and cable.
Figure \ref{fig:architecture} depicts the architecture of the D-profile Ginga DTV Receiver.
Inside the Ginga Middleware, box elements are the non-dashed components, namely, Ginga-NCL and Ginga-CC (Common Core), which are mandatory in any Ginga receiver profile.
Dashed elements are the new components proposed in D-profile receivers, namely, Ginga-HTML5 and Ginga-CC-WebServices.
In the following, we first detail the two mandatory components and then discuss the differences among profiles.

{}

\textbf{Ginga-NCL} subsystem is in charge of receiving and controlling multimedia applications written in NCL.  
NCL \cite{soares_nested_2006} is a declarative multimedia language and its main characteristics are:  
support defining temporal synchronization among media assets and viewer interactions;
layout reuse facilities (<region> and <description>); 
support multi-device presentation; 
support for scripts in the lightweight and embeddable scripting language Lua\footnote{\url{https://www.lua.org}};
and an API for building and modifying applications on-the-fly called NCL editing commands \cite{soares_nested_2006b}.
Ginga-NCL manages applications by grouping them inside a data structure known as a private base. 
The NCL editing commands are divided into three subsets: activation and deactivation of private bases;
life-cycle controls (start, pause, resume, stop);
and updating commands to add and remove elements.

\textbf{Ginga-CC} (Common Core) subsystem is also present and mandatory in profiles. It is a reusable layer responsible for abstracting hardware characteristics.
\textit{Media Players} obtain and render contents delivered in different networks (broadcast and broadband) accessed by a receiver. 
They serve application needs for decoding and presenting content types, including perceptual media content and content that contains declarative (\textit{e.g.} HTML) or imperative code (\textit{e.g.} Lua). 
The \textit{SI} (Service Information) component processes the metadata transmitted by transport stream, such as the SI and PSI (Program Specific Information) tables inside an MPEG-2 TS\footnote{\url{https://www.iso.org/standard/75928}}.
The \textit{DSMCC} component processes the data transmitted by the transport stream, such as applications and stream events.
The \textit{Persistence} component enables broadcasters to store applications and information in non-volatile memory.
In particular, some broadcast commands (\textit{e.g.}, \textit{}{STORED\_AUTOSTART, STORED\_PRESENT, STORED\_REMOVE}) enables to install  applications directly from the DSMCC.
Therefore, this component enables users to browse and launch applications.
It also enables to install and uninstall applications from external memory (USB drive)
or from the Application Store (vendor-specific).

Since its conception, Ginga-NCL supports developers to use different media, \textit{multi-sourced media content}.
For instance, NCL language can handle media content delivered from both broadcast and broadband channels.
However, TV manufactures can create different types of receivers (\textit{e.g.} set-top-box, TV, SmartTV) with different hardware capabilities and prices.
Therefore, the SBTVD Forum defines classes of the receiver with some levels of functionality.
Ginga application developers should target a specific receiver class and signalize it in the delivery channel in order to be processed only by compatible devices. 
We briefly describe the main features of these profiles in what follows\footnote{For more details about the receivers profiles see\cite{abnt_televisao_2018}}.

\begin{description}[leftmargin=5pt]
    \item[A-profile:] \textit{Media Players} support for PNG, JPEG, MNG, AAC, MPEG-1 Audio L2, HTML, Lua scripts and Java code (called Ginga-J\cite{abnt_televisao_2016}).
    \item[B profile:] Added \textit{Media Players} support for MPEG-1 Video
    \item[C-profile:] Added \textit{Media Players} support for H.264 and MP4~(H.264 and AAC). Moreover, it added the \textit{Persistence} component and make the Java code as optional.
    \item[D-profile:] Added \textit{Media Players} support for new immersive (\textit{i.e.} H.265/HEVC, HDR and MPEG-H 3D Audio) and streaming formats (\textit{i.e} HLS and MPEG-DASH), updated Ginga-NCL engine, and added new components Ginga-HTML5 and Ginga-CC-WebServices.
\end{description}

In this paper, we focus on D-profile new features and components.
Therefore, in the next sections, we discuss the new Media Players support~(Section \ref{media-delivery}), the advances in the Ginga-NCL~(Section \ref{ginga-ncl}), and the new components Ginga-HTML5~(Section \ref{Ginga-HTML5}) and Ginga-CC-WebServices~(Section \ref{ginga-cc-ws}).

\section{New \textit{Media Players} Support}\label{media-delivery}

The main updates in \textit{Media Players} are related to better support technologies for video streaming on the Internet and new immersive media.

Regarding video streaming, we cite the support for the emerging \textbf{H.265/HEVC}\cite{itu_recommendation_nodate} video coding standards.
H.265/HEVC is a significant advance in video coding, whereas it consumes up to 50\% less bandwidth than the currently dominant encoding standard H.264/AVC \cite{wiegand_overview_2003}.
Also, the \textit{Media Players} added support for adaptive streaming video formats, such as  
\textbf{MPEG-Dash}\cite{herre_mpeg_h_2015} (mpd files) and \textbf{HLS} (m3u8 files).
This adaptive streaming is based on HTTP and designed to work efficiently over large distributed HTTP networks such as the Internet. 
Its main idea is adjusting in real-time the quality of the media stream accordingly, a user's bandwidth and CPU capacity.
The support of these two transport formats should also support digital copy control through \textbf{DRM}~(Digital Rights Management) according to ISO/IEC 23001-7\footnote{\url{https://www.iso.org/standard/68042.html}}.

Regarding immersive media, we cite the new support for \textbf{HDR and 4K} video formats.
Moreover, \textit{Media Players} also begin to support immersive audio.
The \textbf{MPEG-H 3D Audio}\cite{herre_mpeg_h_2015} standard offers not only immersive 3D audio but also introduces audio objects that can personalize the user experience.
In other words, viewers are capable of individually perceiving single elements of the sound. 
For instance, in a scene with people taking, dialog voices may have different adjustments than background music.

\section{Advances in Ginga-NCL}\label{ginga-ncl}

The C-profile aims to be installed in set-top-box devices for the analog switch-off process.
In order to promote lowering the cost of DTV receivers, SBTVD Forum studied and decided to make Ginga-J optional given its JVM certification costs.
Therefore, one of the main advances for Ginga-NCL engines in D-profile archives fully functional equivalence against the Ginga-J API.
We organize the features of this equivalency in functionalities related to font management, system settings, media seeking and streaming, and other new Lua APIs.

\subsection{Font Support}

The D-profile allows the definition of <font> elements.
These elements should be defined inside the <fontBase> element, at the <head>, and have fontFamily and src attributes.
The src attribute defines a URI for the file containing the source, which must be in a format supported by ABNT NBR 15606-1.
The module thus offers text fonts available for use in descriptors, media node properties, and Lua scripts textit{e.g.}, TTF files). Ginga-NCL also supports new NCL edit commandos to control fonts (\textit{e.g.} addFontBase, addFont).

\subsection{Receiver Network Interfaces}

Regarding network facilities, new NCL settings were added 
in order to provide information about the receiver network interface. The following variables were added to the \texttt{x-ginga-settings} element: 
\texttt{system.macAddress} informs mac address;
\texttt{system.hasActiveNetwork} indicates the existence of any active communication channel interactivity (including LAN interfaces);
\texttt{system.hasNetworkConnectivity }indicates the existence of a channel of active interactivity and with internet connectivity; and
\texttt{system.maxNetworkBitRate} informs the maximum bitrate among
active interactivity channels and with connectivity to
real internet.

The \texttt{}{network.listInterfaces()} is a Lua function that allows access to configuration and status information for each network interface (interactivity channel) available in the receiver. The return of this function is an array with fields related to each network interface (displayName, inetAddress, bcastAddress, subnetMask, inet6Address, hwAddress)
and other enabled features (loopback, pointToPoint, active, supportsMulticast).

\subsection{Media Playback Control}

Ginga-J supported the JMF framework for video control.
Comparing to JMF, Ginga-NCL lacked support for media playback control.
Therefore, new updates in Ginga-NCL added support to handle media seeking and media preparation.

Media seeking is done using the \textbf{\texttt{currentTime}} property in an NCL <media>.
It is valid for continuous media nodes (audio or video), also has its value set by the system, but, in turn, it can have its value modified by attribution actions. 
The \texttt{currentTime} property can either be obtained as a way of monitoring the media object presentation when modified through the start action in an attribution event.
An attribution action in \texttt{currentTime}, with a new playback time (in seconds), leads to re-positioning of the presentation of the media object for the moment expressed in the assigned value.

NCL synchronization is based on events.
An event is an occurrence in time that can be instantaneous
or have a measurable duration.
NCL previously defined three types of events: 
\textit{presentation}, regarding the exhibition of a <media> or a portion of it (<area>);
\textit{selection}, which is defined by selecting a subset of a <media> on display;
\textit{attribution} event, which is defined by assigning a value to a node property.
The D-profile also defines a new \textbf{\textit{preparation}} event,  defined by instantiating and preparing an appropriate media player, including  loading part of media content into the player buffer.
The \texttt{startPrepration} starts the media preparation changing the preparation event state to occurring, while the \texttt{stopPrepration}  ends it.
When a preparation event ends (player media buffer is full), Ginga-NCL triggers the \texttt{onEndPreparation} \texttt{<link>}s.

\subsection{New \textbf{streambuf} media Type}

To render a streamed video in NCL, a developer uses the \texttt{src} attribute of a <media> element to indicate the video streaming URI.
In particular, D-profile allows this URL to refer to adaptive streaming files such as mpd and m3u8 files.
Moreover, D-profile added support for \textbf{\texttt{streambuf://} URI scheme}.
This special scheme enables a developer to directly control the filling of the video buffer.
To do this, the developer requires a Lua script that uses the streambuf class from the event module.

The Lua application inserts data into a streambuf <media> by posting an event in the form:
\texttt{\{class='streambuf', type='write', uri=<uri>, data=buffer\}}.
The "data" is a buffer object that contains the data to be inserted into the streambuf.
This data must conform to the MIME-type declared for the particular NCL media node.
The script obtains the confirmation of the insertion operation when treating the event in 
\texttt{\{class='streambuf', type='write', uri=<uri>, size=number, error=<err\_msg>\}} form.
After a sequence of inserts, the streambuf quickly goes to the FULL state, as there is no consumption by the player instantiated during this preparation. When the streambuf reached FULL during preparation, the Ginga-NCL engine must change its preparation event.

\subsection{Other new Lua APIs}

D-profile defines other new Lua APIs in order to achieve equivalency with the Ginga-J API.
We cite Lua modules for zip compression, charset conversion, transport stream, and bitwise operations. 

The \textbf{zip module} provides the zip object, which is a Lua \texttt{userdata} obtained as a return from an open event of the 'zip' class after their successful treatment.
The zip object offers an API (\texttt{list()}, \texttt{save}) for maintaining entries in a zip file.
Each entry is a record of metadata describing a file or directory contained in the zip file.
For instance, each entry has its path returns by the \texttt{attrPath()}, function.
Other functions related to entries are: \texttt{exists()}, \texttt{remove()}, and \texttt{append()}.

Regarding transport stream, the D-profile profile added,  in event module, the following \textbf{new classes for broadcast information}:
\texttt{pesfilter} enables filtering a particular type of MPEG-2 packetized elementary stream; 
\texttt{psectionfilter} enables filtering a particular type of  MPEG-2 private section; 
\texttt{broadcastfs} enables monitoring changes to an object carousel,such as lost of file, removal, change, or download progress.
Moreover, the new \textbf{"mediakeysession"} supports accessing the handle of the DRM media in the stream.
The Lua code can initiate the procedure when handling events of the encrypted type, which are emitted when starting the
preparation of a media object whose content is protected.
When handling the encrypted event, or proactively, the NCLua application can start creating a set of
keys that the associated media object can later use to decrypt data during presentation.

Other auxiliary modules added were \textbf{charset module} and \textbf{bit32 Lua module}.
The first supports different character encodings when handling strings. \textbf{
charset.convert()} allows string conversion for different character encodings, namely:  'ISO 8859-1', 'UTF-8', 'UTF-16', 'UTF-16BE', and 'UTF-16LE'.
The second offers an API for bitwise operations on 32-bit values.
This module specification is based on a module of the same name from Lua 5.2\footnote{\url{https://www.lua.org/manual/5.2/manual.html\#6.7}}.

The D-profile also added \textbf{ server mode for TCP and UDP} classes in the event Lua module.
This feature is interesting to enable communication among local network devices.
To create a TCP server, the script should post an event with \texttt{tcp} class and \texttt{listen}.
Wheres, to the UDP server, the script should post an event with \texttt{udp} class and \texttt{bind}.

\section{Ginga-HTML5}\label{Ginga-HTML5}

Ginga-HTML5 uses the W3C HTML5 Recommendation \cite{berjon_html5_2014}.
It defines compliance requirements for user-agents and documents that a Ginga-HTML5 engine should comply with.
In particular, a Ginga-HTML5 engine presentation must support HTML5 syntax and XHTML syntax for documents
This way, HTML5 is compatible with the one used in the HbbTV 2.0.

Regarding scripting, the Ginga-HTML5 engine must support scripting as defined in W3C HTML5 Recommendation \cite{berjon_html5_2014} and the ECMA-262-51 scripting language.
The main JavaScript APIs required by a Ginga-HTML5 engine are XMLHttpRequest\footnote{\url{http://www.w3.org/TR/2012/WD-XMLHttpRequest-20121206}}, Web Message\footnote{\url{http://www.w3.org/TR/2012/CR-webmessaging20120501}} and Web Socket\footnote{\url{http://tools.ietf.org/html/rfc6455}}, Web Workers\footnote{\url{http://www.w3.org/TR/2012/CR-workers-20120501}}, Server Sent-Events\footnote{\url{http://www.w3.org/TR/2012/CR-eventsource20121211}}, Web Storage\footnote{\url{http://www.w3.org/TR/2013/REC-webstorage-20130730}},  2D Canvas
and Web Audio API\footnote{\url{https://www.w3.org/TR/2013/WD-webaudio-20131010}.
}.

\section{Ginga-CC WebServices}\label{ginga-cc-ws}

The Ginga-CC-WebServices' main goal is to promote an integrated experience of the broadcasting environment with the home and broadband environments.
To do this, this component provides a set of services implemented in the form of REST APIs\footnote{\url{http://roy.gbiv.com/pubs/dissertation/top.htm}}.
Thus, as indicated in Figure \ref{fig:architecture}, its functionalities can be used by:
1) native Smart TV applications, web-based applications, and Ginga applications in the DTV receiver;
2) devices present in the home environment, allowing, \textit{e.g.}, second-screen scenarios.

The way in which a home device determines the existence of a DTV receiver terminal with support for Ginga-CC-WebServices and obtains the access point to the API is through the SSDP protocol. 
SSDP\footnote{\url{https://tools.ietf.org/html/draft-cai-ssdp-v1-03}} is a text-based protocol that uses the HTTPU\footnote{\url{https://tools.ietf.org/html/draft-goland-http-udp-00}}, which is a variant of HTTP that uses UDP instead of TCP as the transport protocol. 
A Ginga-CC-WebServices server announces its services available by sending  HTTP NOTIFY messages to an IP  multicast address (239.255.255.250 for IPv4 and ff0X::c  for IPv6) and UDP port 1900. 
A home device that wants to discover services available on a network uses the HTTP M-SEARCH method. 
Then the Ginga-CC-WebServices server responds by unicast with a description of the services offered by the DTV receiver.
The HTTP headers regarding this discover and its response are shown in the Listings \ref{list:ssdp-request} and  \ref{list:ssdp-response} bellow.

\begin{lstlisting}[caption={HTTP M-SEARCH headers},frameround=tttt,float=!htp, label={list:ssdp-request}]
M-SEARCH * HTTP/1.1
Host: 239.255.255.250:1900
Man: "ssdp:discover"
ST: urn:sbtvd-org:serviceId:GingaCCWebServices
\end{lstlisting}
\vspace{-3em}

\begin{lstlisting}[caption={HTTP response headers},frameround=tttt,float=!htp, language=xml, label={list:ssdp-response}]
HTTP/1.1 200 OK
Ext: 
Cache-Control: max-age = 5000
ST: urn:schemas-sbtvd-org:service:GingaCCWebServices:1
USN: uuid:<device-UUID>::urn:schemas-sbtvd-org:service:GingaCCWebServices:1
Ginga-CC-Server-BaseURL: <device-IP>:44642
Ginga-CC-Server-SecureBaseURL: <device-IP>:<port>
Ginga-CC-Server-ParingMethods: qrcode, dex
\end{lstlisting}

\vspace{-2em}

\subsection{REST APIs}

All Ginga-CC-WebServices functionalities are exposed through a REST API by an HTTP server implemented as part of the Ginga Common Core. 
Its functionalities include not only functions for accessing content from broadcasting, as well as functions that standardize access to advanced features of the DTV environment (\textit{e.g.},  platform characteristics, supported DRM, media playback capabilities). 
We briefly introduce each API and show its main endpoints in what follows.
In particular, the identifiers in square brackets should be filled in the API request (\textit{e.g.} <host> after the service discover).

\noindent\textbf{API for application authorization:} 
\begin{itemize}[leftmargin=*, topsep=2pt]
    \item \texttt{GET} \url{http(s)://<host>/dtv/authorize}:
    establishes an application link between it and the Ginga-CC-WebServices. 
    The \texttt{pm} param is used to specify the pairing method, whose value may be: \texttt{qrcode} for QRcode paring; \texttt{key} to specify the symmetric key collaborative generation algorithm.
    \item \texttt{GET} \url{http(s)://<host>/dtv/token}: After establishing the initial link and confirming the user's authorization, the application must then obtain the token for secure access to all other APIs.
\end{itemize}
    
\noindent\textbf{API for current DTV service and metadata:} 
\begin{itemize}[leftmargin=*, topsep=2pt]
    \item \texttt{GET} \url{http(s)://<host>/dtv/current-service}: returns information for the selected DTV service;
    \item \texttt{GET} \url{http(s)://<host>/dtv/service-list}: list of available DTV services;
    \item \texttt{GET} \url{http(s)://<host>/dtv/<service-context-id>}: selects a DTV service;
    \item \texttt{GET} \url{http(s)://<host>/dtv/current-service/components}/[<comp-tag>]: returns information about all components (or one given comp-tag param) available in the current service, indicating those that are active;
    \item \texttt{POST} \url{http(s)://<host>/dtv/current-service/components}/<comp-tag>: controls (\textit{e.g.}star, stop) and changes the display (\textit{e.g.} position) of service component.
\end{itemize}

\noindent\textbf{API for handling Ginga environment:} 
\begin{itemize}[leftmargin=*, topsep=2pt]
    \item \texttt{GET} \url{http(s)://<host>/dtv/<service-context-id>/apps/}: lists all Ginga applications available on the service;
    \item \texttt{GET} \url{http(s)://<host>/dtv/<service-context-id>/apps/<appid>}: returns information about one Ginga application;
    \item \texttt{POST} \url{ http(s)://<host>/dtv/current-service/apps/<appid>}: starts or stops a running application.
\end{itemize}
    
\noindent\textbf{API for handling Ginga application properties:}
\begin{itemize}[leftmargin=*, topsep=2pt]
    \item \texttt{GET} \url{http(s)://<host>/dtv/current-service/apps/<appid>/node-properties/<node-id>/[<var-name>]}: returns the value of property defined by an application;
    \item \texttt{GET} \url{ http(s)://<host>/dtv/<service-context-id>/apps/<appid>/persistent/<var-name>/}: returns the value of a property persisted visible to an specific application;
    \item PUT \url{ http(s)://<host>/dtv/<service-context-id>/apps/<appid>/persistent/<var-name>/}: sets value of a property to be persisted visible to a specific application;
    \item \texttt{GET} \url{ http(s)://<host>/dtv/<service-context-id>/ginga/persistent/<var-name>/}: returns the value of property persisted visible to any applications from the same service;
    \item PUT \url{ http(s)://<host>/dtv/<service-context-id>/ginga/persistent/<var-name>/}: sets value of a property to be persisted visible to any applications from the same service.
\end{itemize}

\noindent\textbf{API for handling DSM-CC:}
\begin{itemize}[leftmargin=*, topsep=2pt]
    \item \texttt{GET} \url{http(s)://<host>/dtv/current-service/apps/<appid>/files/}: lists all files from a Ginga application transmitted in the current service;
    \item \texttt{GET} \url{http(s)://<host>/dtv/current-service/dsmcc/files/<component-tag>/<carrousel-id>/[<path>]} lists all files from a given component tag transmitted;
    \item \texttt{GET} \url{http(s)://<host>/dtv/current-service/dsmcc/stream-events/<component-tag>/<carrousel-id>}: subscribes to be notified about stream events;
    \item \texttt{DELETE} \url{http(s)://<host>/dtv/current-service/dsmcc/stream-events/<component-tag>/<carrousel-id>}: unsubscribes a previously registered stream event.
\end{itemize}

\noindent\textbf{API for Ginga-NCL application control} 
\begin{itemize}[leftmargin=*, topsep=2pt]
    \item \texttt{GET} \url{http(s)://<host>/dtv/current-service/apps/<appid>/nodes/[<document-id>]}: lists the nodes (media, contexts and switches) in a running NCL application;
    \item \texttt{POST} \url{http(s)://<host>/dtv/current-service/apps/<appid>/nodes/[<document-id>]}: Allows sending an action (start, stop, pause, resume, abort, select, set, startPreparation, stopPreparation, abortPreparation, and pausePreparation) to a media node, or its interface (<area> or <property>), in a running NCL application;
    \item \texttt{GET} \url{http(s)://<host>/dtv/current-service/ginga/keyset}: lists keys currently reserved for NCL application use;
    \item \texttt{GET} \url{http(s)://<host>/dtv/current-service/ginga/keyset}: sets keys reserved for NCL application use;
    \item \texttt{POST} \url{http(s)://<host>/dtv/current-service/apps/<appid>/nodes/[<document-id>]}: allows sending an \textit{nclEditingCommand} to a running NCL application.
\end{itemize}

\noindent\textbf{API for accessing metadata from the DTV broadcasting:}
\begin{itemize}[leftmargin=*, topsep=2pt]
    \item \texttt{GET} \url{http(s)://<host>/dtv/current-service/info/time|nit|sdt|pat|bat|pmt}: returns MPEG-2 TS table information from the current service; 
    \item \texttt{GET} \url{http(s)://<host>/dtv/<service-context-id>/info/epg}: returns a JSON structure with EPG information from given service;
    \item \texttt{GET} \url{http(s)://<host>/dtv/all\_services/info/epg}: returns a JSON structure with EPG information from all avaliable services.
\end{itemize}

\noindent\textbf{API for accessing multimedia stream from the DTV broadcasting:}
\begin{itemize}[leftmargin=*, topsep=2pt]
    \item \texttt{GET} \url{http(s)://<host>/dtv/current-service/stream/<pid>}: dynamically creates a URI to enable application use  of a multimedia broadcasting stream; this request should return a <handler> identifier;
    \item \texttt{DELETE} \url{http(s)://<host>/dtv/current-service/stream/<handle>}: releases a previous created stream.
\end{itemize}
    
\noindent\textbf{API for accessing information from MediaPlayers:}
\begin{itemize}[leftmargin=*, topsep=2pt]
    \item \texttt{GET} \url{http(s)://<host>/dtv/mediaplayers}: returns the information from all active MediaPlayers, it includes current media, \texttt{currentTime}, supported codecs, etc;
    \item \texttt{GET}  \url{http(s)://<host>/dtv/mediaplayers/<playerid>}: returns the information from one active MediaPlayer;
    \item \texttt{POST}  \url{http(s)://<host>/dtv/mediaplayers/<playerid>}: controls one  active MediaPlayer (\textit{e.g.}, start, stop).
\end{itemize}

\noindent\textbf{API for accessing DTV platform characteristics:}
\begin{itemize}[leftmargin=*, topsep=2pt]
    \item \texttt{GET} \url{http(s)://<host>/dtv/platform-capabilities}: returns  DTV receiver platform's characteristics, capabilities and resources, such as manufacturer, model, colorimetry, physical screen, etc.
    \item \texttt{GET} \url{http(s)://<host>/dtv/drm}: returns supported DRM systems in the platform, this way, an application is able to properly use the decoding and display of a multimedia content.   
\end{itemize}

\noindent\textbf{API for handling native applications using \textit{deep linking}:}
\begin{itemize}[leftmargin=*, topsep=2pt]
    \item \texttt{POST}
    \url{http(s)://<host>/dtv/register-target}: enables a native DTV application to register standardized URL filter (\url{<scheme>:<host>/<path>}) to be accessed later by \textit{deep links}.
    \item \texttt{GET} \url{http(s)://<host>/dtv/target-apps/[target-id]}: lists all (or one by using the target-id param) native applications installed and registered URL filters;
    \item \texttt{POST} \url{http(s)://<host>/dtv/target-apps}: controls (start or stop) a target application or changes its priority (\textit{i.e.} background, foreground);
    \item \texttt{GET} \url{http://<host>/dtv/deep-link/<target-id>}: returns the last \textit{deep link} that activated the registered application. This way, the application can, during its initialization, handle a \textit{deep link} action to be performed.
\end{itemize}

\noindent\textbf{API for controlling access to broadcasting streams:} 
Broadcasters have the exclusive right to provide or revoke access to applications present in the home environment to access their streams.
Therefore, this API should only be valid when called through a client associated to a broadcasted Ginga application (Ginga-HTML5 or Ginga-NCL).
\begin{itemize}[leftmargin=*, topsep=2pt]
    \item \texttt{POST} \url{http(s)://<host>/dtv/bind-context}: binds the currently selected service as a context of service that can be accessed through the Ginga-CC-WebServices APIs;
    \item \texttt{GET} \url{http(s)://<host>/dtv/bind-context}: lists the services that have an identical token as the one sent in the header. After this, the application in home environment is considered authorized to access broadcasting metadata and multimedia streams;
    \item \texttt{DELETE} \url{http(s)://<host>/dtv/bind-context}: revokes a token. 
\end{itemize}

\vspace{-1em}
\section{Use cases}\label{usecases}

New advances in the new D-profile receiver may open many new scenarios for Ginga.

First, given the new \textit{preparation} event, we cite the scenario of \textbf{seamless switch from broadcast to broadband}.
By performing a \texttt{startPreparation} action over a streaming media, an application can smoothly render it, when ready, over the displayed broadcast content.
This way, the viewers' perception of source changes is reduced, giving them a seamless experience.

By tracking users viewing experience, both native SmartTV and Ginga application may perform
\textbf{targeted advertisement/content}.
For instance, after watching a broadcasted content, an application may suggest viewers watch previous or related content in the broadband channel.
Moreover, while watching broadband streaming, an application may suggest viewers to watch live broadcast content related\cite{marina_2019}. 

Another possible scenario is \textbf{multi-device application using Ginga-CC-WebServices}.
By being accessible in the home network, Ginga can interact with other devices to improve the viewer experience.
For instance, second-screen experiences can be facilitated and even goes outside entertainment. It may provide information such as healthcare, education, and wellbeing. 
In particular, in the context of Internet of Things (IoT), \textbf{the DTV receiver can act as an IoT entity}.
For instance, it can connect to house automation  or healthcare devices. 

\section{Final remarks}\label{final-remarks}

This paper introduces the advances and uses cases addressed by the D-profile Ginga DTV receiver.
These specification efforts were carried out in the Brazilian DTV Forum under a collaborative effort led by the Academia. 
However, we must go beyond these new specs and rethink interactivity for future television services.

As future, we may cite other scenarios to be addressed by Ginga advances: supporting Mulsemedia\cite{gheorghita_ghinea_mulsemedia:_2014} application;
supporting 5G broadcasting;
and supporting other new augmenting video content such as scalable video and free viewpoint.

\begin{acks}
We thank TV standardizing committees from the  SBTVD Forum and ITU.
In particular, we thank companies that contribute to the SBTVD forum: Mopa, Mirakulo, Globo, TPV Tech, Samsung, LG, and others.
\end{acks}

\bibliographystyle{ACM-Reference-Format}
\bibliography{refs}

@inproceedings{marina_2019,
author = {Josu\'{e}, Marina and Moreno, Marcelo and Muchaluat-Saade, D\'{e}bora},
title = {Mulsemedia Preparation: A New Event Type for Preparing Media Object Presentation and Sensory Effect Rendering},
year = {2019},
isbn = {9781450362979},
publisher = {Association for Computing Machinery},
address = {New York, NY, USA},
url = {https://doi.org/10.1145/3304109.3306230},
doi = {10.1145/3304109.3306230},
booktitle = {Proceedings of the 10th ACM Multimedia Systems Conference},
pages = {110–120},
numpages = {11},
location = {Amherst, Massachusetts},
series = {MMSys ’19}
}

@article{gheorghita_ghinea_mulsemedia:_2014,
	title = {Mulsemedia: State of the Art, Perspectives, and Challenges},
	volume = {11},
	issn = {1551-6857},
	url = {http://doi.acm.org/10.1145/2617994},
	doi = {10.1145/2617994},
	shorttitle = {Mulsemedia},
	pages = {17:1--17:23},
	number = {1},
	journaltitle = {{ACM} Transactions on Multimedia Computing, Communications, and Applications},
	author = {{Gheorghita Ghinea} and Timmerer, Christian and Lin, Weisi and Gulliver, Stephen R.},
	urldate = {2014-10-21},
	year = {2014}
}

@article{soares_nested_2006,
	title = {Nested Context Model 3.0 Part 6 – {NCL} (Nested Context Language) Main Profile},
	issn = {0103-9741},
	year = {2006},
	url = {ftp://ftp.inf.puc-rio.br/pub/docs/techreports/06_15_soares.pdf},
	journaltitle = {Monographs in Computer Science {PUC}-Rio Inf {MCC}15/06},
	author = {Soares, Luiz Fernando Gomes and Rodrigues, Rogério Ferreira and Costa, Romualdo Monteiro de Resende},
}

@article{soares_nested_2006b,
	title = {Nested Context Language 3.0 Part 9 – {NCL} Live Editing Commands},
	issn = {0103-9741},
	url = {ftp://ftp.inf.puc-rio.br/pub/docs/techreports/06_36_soares.pdf},
	shorttitle = {Nested Context Language 3.0},
	journaltitle = {Monographs in Computer Science {PUC}-Rio Inf {MCC}36/06},
	author = {Soares, Luiz Fernando Gomes and Rodrigues, Rogério Ferreira and Costa, Romualdo Rezende and Moreno, Marcio Ferreira},
	year = {2006}
}

@online{berjon_html5_2014,
	title = {{HTML}5 W3C Candidate Recommendation},
	url = {https://www.w3.org/TR/2014/REC-html5-20141028},
	titleaddon = {World Wide Web Consortium},
	author = {Berjon, Robin and Faulkner, Steve and Leithead, Travis and Navara, Erika Doyle and O’Connor, Edward and Pfeiffer, Silvia and Hickson, Ian},
	year = {2014}
}

@article{herre_mpeg_h_2015,
	title = {{MPEG}-H 3D Audio—The New Standard for Coding of Immersive Spatial Audio},
	volume = {9},
	issn = {1941-0484},
	doi = {10.1109/JSTSP.2015.2411578},
	pages = {770--779},
	number = {5},
	journaltitle = {{IEEE} Journal of Selected Topics in Signal Processing},
	author = {Herre, Jürgen and Hilpert, Johannes and Kuntz, Achim and Plogsties, Jan},
	year = {2015},
	note = {Conference Name: {IEEE} Journal of Selected Topics in Signal Processing}
}

@article{wiegand_overview_2003,
	title = {Overview of the H.264/{AVC} video coding standard},
	volume = {13},
	issn = {1558-2205},
	doi = {10.1109/TCSVT.2003.815165},
	pages = {560--576},
	number = {7},
	journaltitle = {{IEEE} Transactions on Circuits and Systems for Video Technology},
	author = {Wiegand, T. and Sullivan, G.J. and Bjontegaard, G. and Luthra, A.},
	year = {2003},
	note = {Conference Name: {IEEE} Transactions on Circuits and Systems for Video Technology}
}

@report{itu_recommendation_nodate,
	title = {Recommendation H.265: High efficiency video coding},
	url = {https://www.itu.int/rec/T-REC-H.265-201911-I/en},
	author = {{ITU}},
	year = {2019}
}

@report{abnt_televisao_2018,
	title = {Televisão digital terrestre — Receptores},
	url = {https://www.abntcatalogo.com.br/norma.aspx?ID=406737},
	author = {{ABNT}},
	urldate = {2020-03-12},
	year = {2018}
}

@report{abnt_televisao_2016,
	title = {Televisão digital terrestre — Codificação de dados e especificações de transmissão para radiodifusão digital Parte 4: Ginga-J - Ambiente para a execução de aplicações procedurais},
	url = {https://www.abntcatalogo.com.br/norma.aspx?ID=351841},
	author = {{ABNT}},
	urldate = {2020-03-12},
	year = {2016}
}

@report{itu2014h761,
    author={ITU},
    institution={{ITU}},
    title={Recommendation H.761: Nested Context Language (NCL) and Ginga-NCL for IPTV Services},
    address={Geneva, Switzerland},
    publisher={ITU-T},
    url={https://www.itu.int/rec/T-REC-H.761},
    year={2014},
}

\end{document}